# Origin of topologically trivial states and topological phase transitions in low-buckled plumbene

Yue Li,[1] Jiayong Zhang,[2,1] Bao Zhao,[3,1] Yang Xue,[1] and Zhongqin Yang[1,4*]

1. State Key Laboratory of Surface Physics and Key Laboratory for Computational Physical Sciences (MOE) and Department of Physics, Fudan University, Shanghai 200433, China

2. $\Psi_{usts}$ Institute, Jiangsu Key Laboratory of Micro and Nano Heat Fluid Flow Technology and Energy Application, and School of Mathematics and Physics, Suzhou University of Science and Technology, Suzhou, Jiangsu 215009, China

3. School of Physics Science and Information Technology, Shandong Key Laboratory of Optical Communication Science and Technology, Liaocheng University, Liaocheng 252059, China

4. Collaborative Innovation Center of Advanced Microstructures, Fudan University, Shanghai 200433, China

**ABSTRACT**: Combining tight-binding (TB) models with first-principles calculations, we investigate electronic and topological properties of plumbene. Different from the other two-dimensional (2D) topologically nontrivial insulators in group IVA (from graphene to stanene), low-buckled plumbene is a topologically trivial insulator. The plumbene without spin-orbit coupling exhibits simultaneously two kinds of degeneracies, i.e., quadratic non-Dirac and linear Dirac band dispersions around the Γ and K/K' points, respectively. Our TB model calculations show that it is the coupling between the two topological states around the Γ and K/K' points that triggers the global topologically trivial property of plumbene. Quantum anomalous Hall effects with Chern numbers of 2 or -2 can be, however, achieved after an exchange field is introduced. When the plumbene is functionalized with ethynyl (PbC$_2$H), quantum spin Hall effects appear due to the breaking of the coupling effect of the local topological states.

**I. Introduction:**

Recently, quantum spin Hall (QSH) insulators carried out in two-dimensional (2D) materials have attracted intense interest in condensed matter physics and material science, whose spin-momentum locked dissipationless gapless edge states inside the bulk energy gaps are protected by time-reversal symmetry (TRS) [1] and favorable for promoting the development of low-power-consumption microelectronic devices. The landmark concept of the QSH effect was first put forth in graphene by Kane and Mele in 2005 [2,3]. Unfortunately, the spin-orbit coupling (SOC) of graphene is very weak that the band gap opened up by the SOC at the degenerate Dirac point turns out to be on the order of $10^{-5}$ eV [4], inferring that the QSH effect in graphene can be only realized at an unrealistically low temperature. Subsequently, the other 2D group IVA film materials of silicene, germanene, and stanene with low-buckled honeycomb structures were synthesized in experiments [5-13] and also reported to be topological insulators hosting the QSH effect [14-17] due to the similar Dirac physics as in graphene. The case for the plumbene composed by the last element Pb in group IVA is, however, very different. The atomic Pb film with low-buckled configuration has been theoretically predicted to be as a topologically trivial insulator in Refs. [18-20] rather than QSH insulators found in other group IVA atomic films. Some researchers have been performed to tune the low-buckled plumbene from a topologically trivial insulator to a topologically nontrivial insulator, by such as chemical or electron doping [18,20-23]. The reason why the topological behavior of the low-buckled plumbene is different from those of other group IVA atomic film is, however, not clear yet.

In this work, based on tight-binding (TB) models and first-principles calculations, we demonstrate the novel band dispersions and origin of the topologically trivial properties of the low-buckled plumbene. A TB Hamiltonian with sixteen bands based on Pb $s$, $p_x$, $p_y$, and $p_z$ orbitals in a honeycomb lattice is built. Besides the linear Dirac bands obtained at the K and K' points, similar to the cases in other group IVA monolayers, the TB results give a quadratic non-Dirac bands with two-fold degeneracy at the Γ point around the Fermi level ($E_F$). The two-fold degenerate bands composed of Pb $p_x$ and $p_y$ orbitals at the Γ point comes from the $C_3$ symmetry of the honeycomb atomic lattice, which are unique in plumbene and do not appear around the $E_F$ for the other group IVA monolayers. Through the calculations of Berry curvatures for the Hamiltonian in the single spin subspace, we find that both the local states around the Γ and K/K' points are topologically nontrivial. The coupling between the two topological states around the Γ and K/K' points, however, gives rise to the global topologically trivial property of plumbene. A quantum anomalous Hall (QAH) state with a Chern number of 2 or -2 is predicted in the system if an exchange interaction is included. A giant band-gap QSH insulator is achieved in the plumbene film if it is functionalized with ethynyl ($C_2H$) due to the breaking of the coupling effect of the two topological states. Our work unveils the reason why the electronic structure of the low-buckled plumbene is topologically trivial instead of nontrivial, different from the cases of the other 2D group IVA materials. Our work also provides a new theoretical path to search new 2D topological materials.

**II. Models and methods**

Pb atoms in plumbene are arranged in a honeycomb lattice, which could have three stable structures, namely planar, low-buckled, and high-buckled configurations [22,24]. Similar to Si and Ge atoms etc.[14], Pb atoms also prefer $sp^3$ bonding besides the $sp^2$ bonding. Hence the two buckled configurations are more stable than the planar one. For plumbene and stanene, the most stable structures were reported to be high-buckled [22,24], different from those of the silicene and germanene. In this work, we focus on the low-buckled configuration, the meta-stable structure, for plumbene since the low-buckled structure (Fig. 1 (a) and (b)) owns the interesting Dirac cones around the K and K' points while an exotic normal insulator is predicted in this material [18-20].

The geometry optimization and electric structure calculations of the plumbene are performed with projected augmented wave (PAW) [25] formalism based on ab initio density-functional theory (DFT), as implemented in the Vienna ab initio simulation package (VASP) [26]. The Perdew-Burke-Ernzerhof generalized-gradient approximation (GGA-PBE) is employed to describe the exchange and correlation functional [27]. The plane-wave cutoff energy is set to be 400 eV and a vacuum space 20 Å is set to avoid the interactions between the two adjacent slabs. The convergence criterion for the total energy is $10^{-6}$ eV. All the Pb atoms in the low-buckled plumbene are allowed to relax until the Hellmann-Feynman force on each atom is smaller than 0.01 eV/Å. The gamma central Monkhorst-Pack grids of 9×9×1 are adopted.

## III. Results and discussion

**A. Electronic structures and topological properties of plumbene**

The optimized geometry of the low-buckled structure is shown in Fig. 1 (a) and (b), with the obtained lattice constant of 4.934 Å and a height difference ($h$) of 0.99 Å, consistent with the results reported by others [21,22]. The band structures of the low-buckled plumbene obtained from the first-principles calculations are illustrated in Fig. 1 (c) and (d). In the absence of SOC, linear Dirac band dispersion composed of $p_z$ orbital is found around the K/K' points in the vicinity of $E_F$ (Fig. 1(c)), similar to the cases of other group IVA monolayers. As a representative, the silicene bands without SOC are displayed in Fig. 1(e). Previous studies showed that band gaps can be opened by SOC around the Dirac point in graphene, silicene to stanene (low-buckled structure) and QSH effects can come out in these systems [14-17]. Due to the C$_3$ symmetry of the honeycomb structure, another set of two-fold degenerate bands happens at the Γ point in silicene (Fig. 1(e)) below the E$_F$, composed of Si $p_x$ and $p_y$ orbitals. The dispersion of these $p_x$ and $p_y$ orbitals belongs to quadratic non-Dirac since the Γ point is a time-reversal invariant point and the linear k term in the energy eigenvalue at such point is forbidden [28]. These two-fold $p_x$/$p_y$ bands move up to the E$_F$ in the case of plumbene (Fig. 1(c)). Due to the conservation of valence electrons in the systems, the $s$ antibonding state ($s^-$) becomes an occupied state in the plumbene (Fig. 1(c)), while it is not in the case of the silicene (Fig. 1(e)). Note that the major contribution of the $s^-$ state is located around -5 eV, not displayed in Fig. 1(c). Therefore, for the plumbene, there are two types of bands in the vicinity of the E$_F$: quadratic non-Dirac bands around the Γ point and linear Dirac bands around the K/K' points, hinting the unique electronic transport behaviors of the plumbene, compared to other group IVA monolayers.

When the SOC of Pb atoms is taken into account, the degeneracies at the Γ and K/K' points are both lifted and a global band gap of 0.44 eV is opened (Fig. 1(d)). Its topology can be identified by the edge states. The calculated bands of a plumbene ribbon system based on the maximally localized Wannier functions (MLWFs) are illustrated in Fig. 1(f), indicating explicitly the topologically trivial edge states within the band gap around the E$_F$. They connect, however, only the valence bands instead of conduction and valence bands. Therefore, the low-buckled plumbene is a topologically trivial insulator rather than a QSH insulator owned by other group IVA monolayers, consistent with the results of topological invariant Z$_2$=0 obtained in Ref. [21]. As mentioned above, a band inversion between $s$ antibonding state ($s^-$) and $p_x$/$p_y$ bonding states ($p_x$/$p_y^+$) occurs in the low-buckled plumbene with respect to the other group IVA monolayers, which can cause the topological phase transition in the plumbene from the point of view of the parity changes [29].Thus, from the first-principles calculations combined with the MLWFs, we obtain the exotic band dispersion and topologically trivial property of the plumbene. To deeply understand the origin of the topologically trivial states in the plumbene with a unique band dispersion around the $E_F$, a TB model with multiple orbitals is constructed.

**B. Tight-binding model and results**

Since the orbital-projected band structures displayed in Fig. 1(c) and (d) show the contributions from all of $s$ and $p$ ($p_x$, $p_y$, and $p_z$) orbitals around the $E_F$, we derive a TB Hamiltonian for the plumbene in the basis of four orbitals ($s$, $p_x$, $p_y$, and $p_z$) of Pb atoms. There are two Pb atoms in one unit cell as shown in Fig. 1(a). The basic functions in each inequivalent site are taken as $|s_\uparrow\rangle$, $|p_{x\uparrow}\rangle$, $|p_{y\uparrow}\rangle$, $|p_{z\uparrow}\rangle$; $|s_\downarrow\rangle$, $|p_{x\downarrow}\rangle$, $|p_{y\downarrow}\rangle$, $|p_{z\downarrow}\rangle$. Hence, the total Hamiltonian of this honeycomb lattice can be written as

$$H(k) = H_{hopp}(k) + H_{soc} + H_M, \tag{1}$$

where each term is given by

$$H_{hopp}(k) = \sum_{<ij>,\alpha,\beta;\sigma,\sigma'} c^\dagger_{\alpha\sigma i} t_{\alpha\sigma i;\beta,\sigma' j} c_{\beta\sigma' j}, \tag{2}$$

$$H_{soc} = \lambda \mathrm{L} \cdot \mathrm{S}, \tag{3}$$

$$H_M = M \sum_{i;\alpha;\sigma,\sigma'} c^\dagger_{\alpha\sigma i} s^z_{\sigma\sigma'} c_{\alpha\sigma' i}. \tag{4}$$

The $c^\dagger_{\alpha\sigma i}$ ($c_{\alpha\sigma i}$) represents the creation (annihilation) operators for an electron with spin $\sigma$ and orbital $\alpha$ on site $i$. $t_{\alpha\sigma i;\beta,\sigma' j}$ is the nearest-neighbor (NN) hopping integrals parameter, which can be given by the Slater-Koster integrals [30,31]. < > runs over all the NN hopping sites. The first term $H_{hopp}(k)$ indicates the NN hopping term, with $k = (k_x, k_y)$. The second term $H_{soc}$ represents the on-site SOC Hamiltonian, where $\lambda$ is the atomic SOC strength. And the third term $H_M$ describes magnetic exchange field term, in which $M$ parameter is the external exchange field strength and $s$ is the spin Pauli matrices.

The detailed derivations of $H_{hopp}(k)$, $H_{soc}$, and $H_M$ from the Slater-Koster integrals [30,31] are specified in Supporting Note 1 of the Supplemental Material. In total, the TB model leaves us with seven parameters of $e_s$, $e_p$, $V_{ss\sigma}$, $V_{sp\sigma}$, $V_{pp\sigma}$, $V_{pp\pi}$, and $\lambda$ to be fitted with the first-principles results of the plumbene. The fitted band structures without and with SOC from the built TB model are shown in Fig. 2(a) and (b), respectively, with fitted parameters $e_s = -7.22$ eV, $e_p = 0.84$ eV, $V_{ss\sigma} = -0.85$ eV, $V_{sp\sigma} = 1.4$ eV, $V_{pp\sigma} = 1.475$ eV, $V_{pp\pi} = -0.7$ eV, and $\lambda = 0.46$ eV. Consistent with the first-principles bands (Fig. 1(c) and (d)), in the absence of SOC, there are not only linear Dirac band dispersions at the K/K' points around the $E_F$ but also quadratic non-Dirac band dispersions at the $\Gamma$ point simultaneously. When SOC is taken into account, the degeneracies at the $\Gamma$ and K/K' points are both lifted and band gaps are opened around these points due to the finite SOC strength. In comparison to Fig. 1(d), a global gap of 0.44 eV can be opened with the atomic SOC strength $\lambda = 0.46$ eV. Obviously, the electronic structure of the low-buckled plumbene can be described well through the TB method built.

## C. Coupling effect of topological states in plumbene

The Berry curvatures are calculated in terms of the 16-band Hamiltonian obtained from the TB model to understand the topological behavior of the low-buckled plumbene. The Berry curvatures are calculated with Kubo formula [32-34],

$$\Omega_{\alpha\beta}(\mathbf{k}) = \sum_n f_n \Omega_{n,\alpha\beta}(\mathbf{k}),$$
$$\Omega_{n,\alpha\beta}(\mathbf{k}) = -2\mathrm{Im}\sum_{m\neq n}\frac{\hbar^2\langle\psi_{n\mathbf{k}}|v_\alpha|\psi_{m\mathbf{k}}\rangle\langle\psi_{m\mathbf{k}}|v_\beta|\psi_{n\mathbf{k}}\rangle}{(E_m(k)-E_n(k))^2}, \quad (5)$$

where the Greek letters $\alpha$, $\beta$ indicate Cartesian coordinates, $f_n$ is the Fermi-Dirac distribution function, $v_\alpha$ and $v_\beta$ are the Cartesian velocity operators, and the summation is over all of the occupied states. The Chern number $C$ is obtained by integrating the $\Omega(\mathbf{k})$ over the first Brillouin zone (BZ), $C = \frac{1}{2\pi}\sum_n \int_{BZ} d^2k \Omega_n$ [35].

As discussed above, the states around the $E_F$ are contributed mainly by the Pb $p_x/p_y$ and $p_z$ orbitals. The interactions between these multiple $p$ ($p_x$-$p_z$ or $p_y$-$p_z$) orbitals can give rise to non-diagonal terms of spin-up and spin-down states. Thus, the spin is no longer a good quantum number, causing that the spin Chern number is inappropriate to characterize the topological properties in this TB model. To avoid this problem, we apply a large exchange field $M = 5.5$ eV to separate the spin-up and spin-down bands with distinguished energy gaps, as shown in Fig. 3(a), so that the coupling of spins becomes weak and can be ignored. By setting the $E_F$ inside the band gap of $p_x/p_y$ orbitals, opened by SOC, we calculate the corresponding Berry curvatures for the valence bands for each spin channel, as shown in Fig. 3(b-d).

Fig. 3(b) illustrates the Berry curvatures (pink dots) and the magnified surrounding bands (green dots) for the spin-up subspace Hamiltonian when the $E_F$ is tuned into the band gaps of the $p$ ($p_x/p_y$ and $p_z$) orbitals, indicated by the pink dashed line in Fig. 3(b). Two very sharp Berry curvature peaks appear around the K and K' points, while two fat Berry curvature peaks present around the $\Gamma$ point in Fig. 3(c). We label $C_\Gamma$, $C_K$, and $C_{K'}$ as the integrating values of the Berry curvatures around the $\Gamma$, K, and K' points, respectively. In the spin-up subspace, we obtain $C_{\Gamma\uparrow} = 1$, $C_{K\uparrow} = 0.5$, and $C_{K'\uparrow} = 0.5$ and consequently the Chern number of the spin-up subspace is $C_\uparrow = C_{\Gamma\uparrow} + C_{K\uparrow} + C_{K'\uparrow} = 2$. The distribution of the Berry curvatures in the 2D momentum space is also shown in Fig. 3(c), from which the positive contributions at the K/K' points as well as at the places surrounding the Γ point can be clearly seen, indicating the bands around the K/K' and Γ points both contribute to the Chern number, consistent with the local Chern numbers obtained at the K/K' and Γ points. The distribution of the Berry curvatures displayed in Fig. 3(c) shows evident $C_3$ symmetry due to the honeycomb lattice of the plumbene (Fig. 1(a)).

Similarly, the spin-down case of the $p$ ($p_x/p_y$ and $p_z$) orbitals is illustrated in Fig. 3(d). The Berry curvature peaks hold the opposite sign, as compared to the spin-up situation; namely, $C_{\Gamma\downarrow} = -1, C_{K\downarrow} = -0.5, C_{K'\downarrow} = -0.5$, and $C_\downarrow = C_{\Gamma\downarrow} + C_{K\downarrow} + C_{K'\downarrow} = -2$. These results give the spin-up and spin-down Hamiltonians belong to the $C_\uparrow = 2$ and $C_\downarrow = -2$ QAH insulators, respectively. Therefore, when a large spin polarization is introduced in the plumbene, QAH states (with C=2 or -2) can be achieved in the spin-up or spin-down channels of the plumbene. Since the Berry curvatures at the Γ and K/K' points have the same sign for a certain spin channel, the coupling effect of the topological states around the Γ and K/K' point in plumbene can be called 'constructive interference', different from the 'destructive interference' effect reported in Ref. [36].

If the spin polarization is removed from the plumbene, we can expect that there are two pairs of chiral spin currents flowing along the sample edges. Since the TRS only exists within one pair of the chiral spin current, backscattering can occur between the two pairs of the edge states, leading to the destruction of the two pairs of topological protected edge states with each other. Thus, a global $Z_2 = 0$ topologically trivial state is finally achieved in the system. In short, the global topologically trivial state of the low-buckled plumbene is resulting from the 'constructive' coupling effect between the two local topologically nontrivial states around the Γ point with $p_x/p_y$ orbitals and K/K' points with $p_z$ orbital. The unavoidable interaction between the two pairs of the topological protected edge states leads to the breakdown of the QSH effect in the system. If there are odd number (such as 3 or 5) edge states, the QSH state will survive in the system. QAH effects can always be obtained if a large spin polarization is introduced in the system due to the 'constructive' coupling effect of the local topological states at the $\Gamma$ and K/K' points. No such coupling effect of topological states exists in the other 2D group IVA monolayer materials, whose topology is determined only by the Dirac bands at the K/K' points. This is the reason why the topological behaviors of the low-buckled plumbene are so distinguished from those of the other group IVA monolayers.

**D. Topological phase transitions tuned based on the coupling effect**

Since the topologically trivial property of the low-buckled plumbene results from the coupling effect between two topologically nontrivial states around the $\Gamma$ and K/K' points, one direct strategy to realize the topologically nontrivial state in the plumbene is to break the coupling effect by shifting one of the band degeneracies around the $\Gamma$ and K/K' points away from the $E_F$. On the one hand, as introduced above, from the plumbene to the other 2D group IVA monolayer materials, the energy position of the quadratic non-Dirac bands decreases from at the $E_F$ to below the $E_F$. Therefore, there is no coupling effect in the other 2D group IVA monolayer materials, which can rationalize well the topologically nontrivial states predicted in these materials [14-17]. On the other hand, the position of the linear Dirac bands around the K/K' points can also be moved away from the $E_F$. As an example, we propose a way to carry out a topologically nontrivial state through this path. When the low-buckled plumbene is functionalized with ethynyl (forming PbC$_2$H), the $p_z$ orbital of Pb is saturated by ethynyl, which may cause the linear Dirac bands around the K/K' points to move away from the $E_F$. Consequently, the coupling effect at the $E_F$ can be broken.

The optimized structure of PbC$_2$H obtained from the DFT calculations is shown in Fig. 4(a) and (b), with the lattice constant $a = 5.256$ Å and the vertical distance of the two non-coplanar Pb atoms $h = 0.574$ Å. The lattice constant of PbC$_2$H becomes larger by 6.5% compared to the pristine low-buckled plumbene film primarily due to the interaction between ethynyl and Pb atoms. As illustrate in Fig. 4(c), in the absence of SOC, there is only the quadratic non-Dirac bands around the Γ point (composed of $p_x$ and $p_y$ orbitals of Pb) in the vicinity of the E$_F$, as expected. Large band gaps are opened at the K/K' points due to the Pb $p_z$ orbital saturated by the added ethynyl. Hence, the coupling effect between the two topological states around the Γ and K/K' points is broken. And the globally topological behavior of the system is determined only by the bands around the Γ point. After SOC is taken into account, a QSH effect with a band gap of 0.912 eV will appear in the system. The topological mechanism around the Γ point with two-fold degenerate quadratic energy dispersions composed of $p_x$ and $p_y$ orbitals was investigated deeply in Refs. [37-40], which is totally different from the typical Bernerg-Hughes-Zhang (BHZ) model for HgTe/CdTe quantum wells with band inversion [41] and Kane-Mele model for linear Dirac band dispersion [2,3] for carrying out the topological states. The two-fold degenerate bands of $p_x$ and $p_y$ orbitals around the E$_F$ is guaranteed by the C$_3$ symmetry of the lattice.

The topology of the plumbene decorated with ethynyl is identified in quantitatively by the calculations of the edge states and topological invariant. The bands of a PbC$_2$H ribbon is calculated based on the Wannierized Hamiltonian, illustrated in Fig. 4(e), revealing explicitly that the gapless edge states (red lines) exist inside the bulk energy gap of PbC$_2$H. And the topological invariant $Z_2 = 1$ is obtained by using the Wannier charge center (WCC) method [42]. The topologically nontrivial edge states and the topological invariant $Z_2 = 1$ both guarantee the QSH effect hosted in PbC$_2$H. Thus, we successfully tune the topologically trivial insulator of the low-buckled plumbene to a topologically nontrivial insulator, based on the understanding to the origin of the topologically trivial states of the plumbene.

The structural stability of PbC$_2$H is also investigated. The saturation of dangling bonds of $p_z$ orbital of the plumbene by ethynyl is found making the system more stable. The total energy of PbC$_2$H as a function of the lattice constant is displayed in Fig. S1, indicating that the low-buckled structure is more stable than the high-buckled one and becomes the most stable geometry. It can be understood intuitively that the saturation is favorable to the $sp^2$ bonding instead of $sp^3$ bonding. To investigate the dynamical stability of PbC$_2$H, the phonon dispersion of PbC$_2$H is calculated [43]. No imaginary frequency is found in the phonon dispersion (Fig. 4(f)). These results indicate the structural stability and accessibility of the system in experiments.

## IV. Conclusions

We investigate why the topological behavior of the low-buckled plumbene is different from the other group IVA monolayers. In the absence of SOC, besides the linear Dirac band dispersions around the K/K' points at the E$_F$, the Pb $p_x$ and $p_y$ orbitals also form the two-fold degenerate non-Dirac bands around the Γ point at the E$_F$. With the 16-band TB model, we find the 'constructive' coupling effect of the local topologically nontrivial states around the Γ and K/K' points leads to the two pairs of the edge states carrying spin current and finally causes the breakdown of the QSH effects in the plumbene. QAH effects with the Chern number of 2 or -2 can appear in the system if an exchange interaction is introduced. No such coupling effect, however, exists in the other group IVA monolayers, so all of them host the QSH effect. The QSH effect can be resumed in the plumbene decorated with ethynyl due to the breaking of the coupling effect of the topological states at the $E_F$. Our work rationalizes well the topological behaviors of the plumbene and importantly provides one new mechanism to find or design various topological states in materials.

## ACKNOWLEDGMENTS

This work was supported by the National Natural Science Foundation of China under Grant Nos. 11574051, 11874117, 11604134, and 11747137, and the Natural Science Foundation of Jiangsu Province (China) with Grants No. BK20170376. All the calculations were performed at the High Performance Computational Center (HPCC) of the Department of Physics at Fudan University.

## AUTHOR INFORMATION

**Corresponding Author**

*E-mail: zyang@fudan.edu.cn

**Figures and captions**

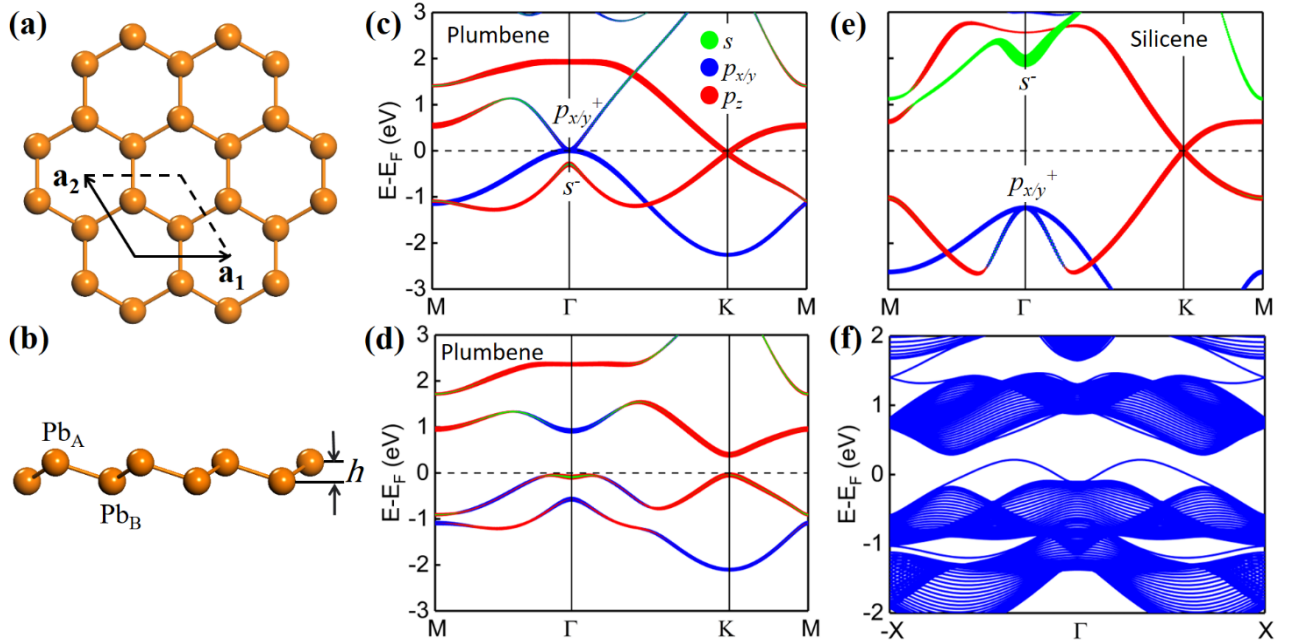

**Figure 1.** (a) and (b) The geometry structures of the low-buckled plumbene from the top and side views, respectively. The black solid and dotted lines indicate the unit cell. $Pb_A$ and $Pb_B$ sublattices are not coplanar. The height $h$ is defined as the vertical distance between the $Pb_A$ and $Pb_B$ planes. (c) and (d) give the orbital-projected band structures of the plumbene without and with SOC, respectively. The sizes of the dots are proportional to the contribution of the corresponding orbitals. (e) gives the orbital-projected band structure of silicene without SOC. (f) The calculated bands of the plumbene ribbon based on the MLWFs.

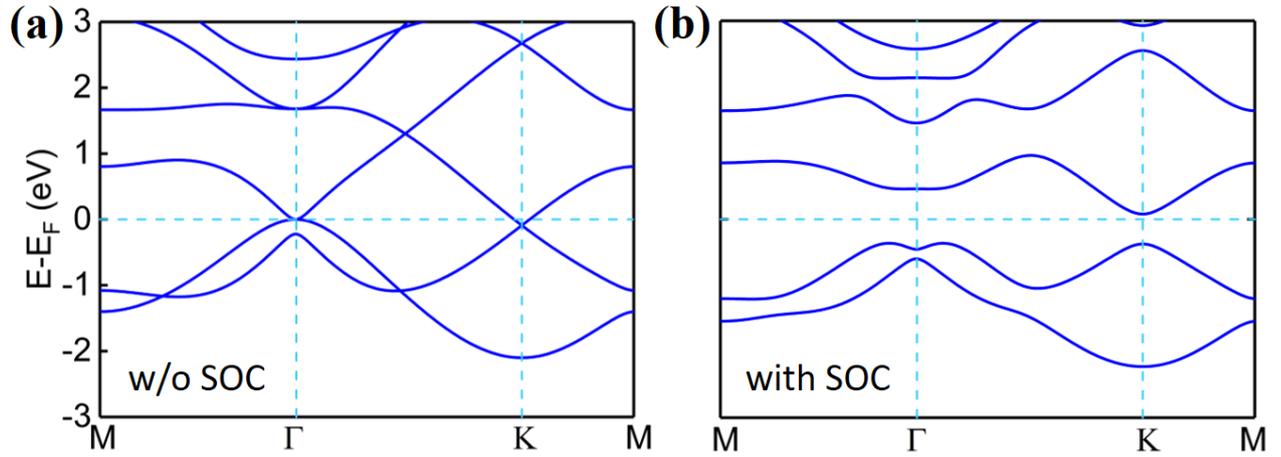

**Figure 2.** Band structures of the plumbene obtained from the TB model with four orbitals $(s, p_x, p_y, p_z)$ per nonequivalent lattice site. The parameters adopted are $e_s = -7.22$ eV, $e_p = 0.84$ eV, $V_{ss\sigma} = -0.85$ eV, $V_{sp\sigma} = 1.4$ eV, $V_{pp\sigma} = 1.475$ eV, $V_{pp\pi} = -0.7$ eV. (a) The SOC is not considered ($\lambda = 0.0$ eV). (b) The SOC of $\lambda = 0.46$ eV is considered.

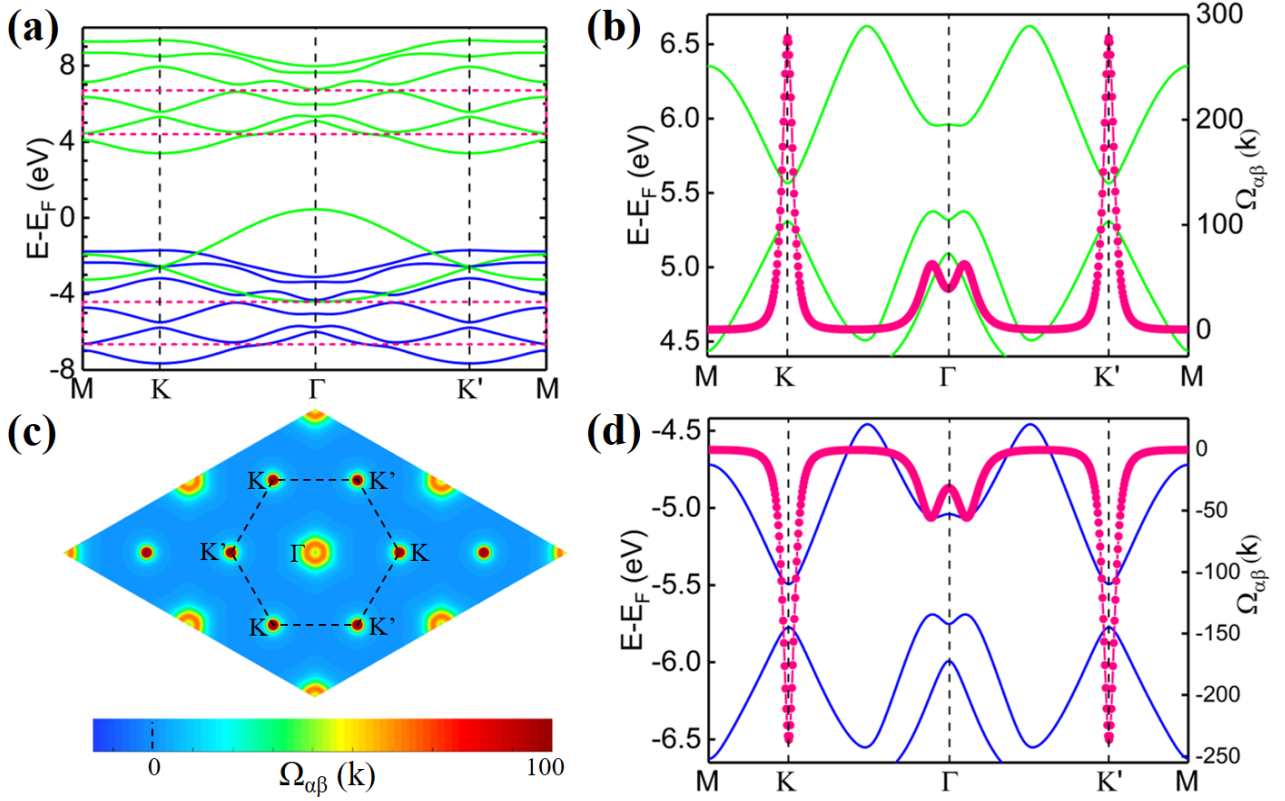

**Figure 3.** (a) Band structure with the same parameters as in Fig. 2(b) except that an exchange interaction ($M = 5.5$ eV) is considered. The green and blue curves denote the spin-up and spin-down bands, respectively. (b) The Berry curvatures (pink dots) and the corresponding bands (green curves) for the spin-up subspace of the TB Hamiltonian. The bands are the magnified plot framed by the pink dashed rectangle around 5.5 eV in (a). In the calculations of the Berry curvatures, the $E_F$ is set within the band gap of Pb $p_x/p_y$ orbitals, as shown by the pink dashed line. (c) The distribution of the Berry curvatures in the 2D momentum space for the spin-up subspace of the TB Hamiltonian. (d) is the same as in (b), except for the spin-down subspace of the TB Hamiltonian. The bands are the magnified plot framed by the pink dashed rectangle around -5.5 eV in (a).

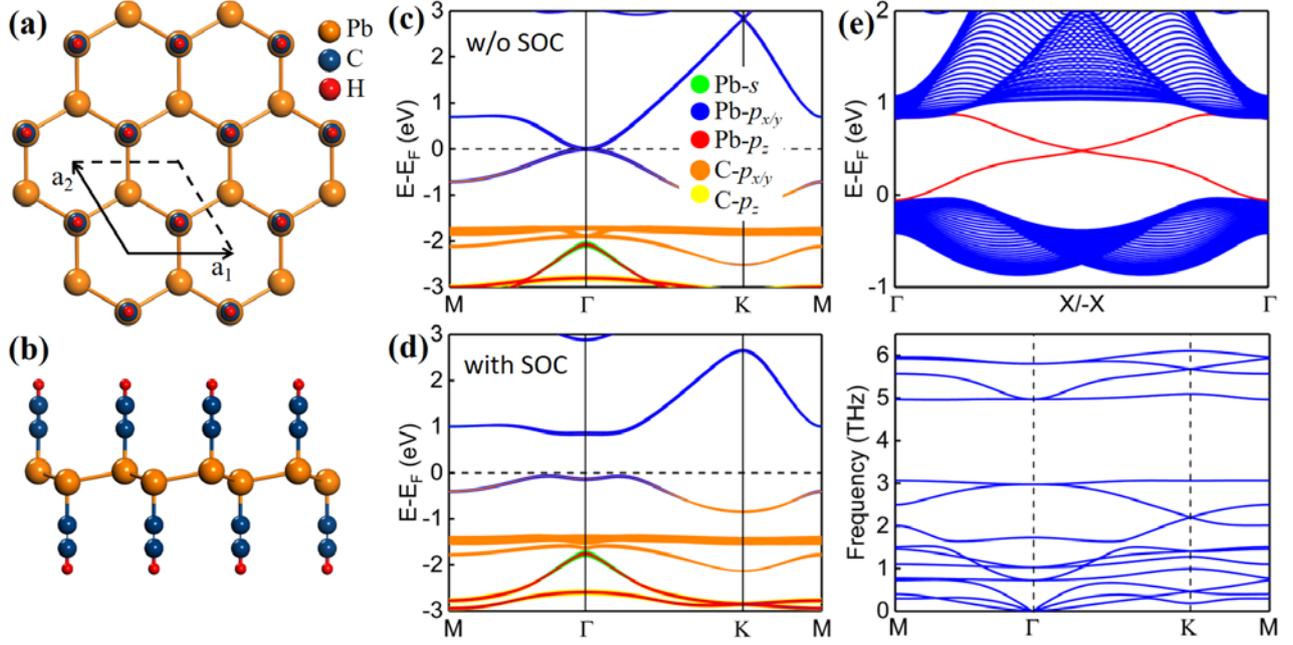

**Figure 4.** (a) and (b) give the geometry structures of PbC$_2$H from the top and side views, respectively. The black solid and dotted lines indicate the unit cell. (c) and (d) are orbital-projected band structures of PbC$_2$H without and with SOC, respectively. The green, blue, red, orange, and yellow circles represent $s$, $p_x/p_y$, $p_z$ orbitals of Pb, and $p_x/p_y$, $p_z$ orbitals of C, respectively. The sizes of the circles are proportional to the contribution of the corresponding orbitals. (e) The calculated bands of PbC$_2$H ribbon systems based on the Wannierized Hamiltonian. (f) Phonon band dispersions of PbC$_2$H.